# Modal Coupling of Single Photon Emitters Within Nanofibre Waveguides

*Michele Gaio[◆], Maria Moffa[○], Marta Castro-Lopez[◆], Dario Pisignano[○,■]\*, Andrea Camposeo[○]\*, and Riccardo Sapienza[◆]\**

◆ Department of Physics, King's College London, Strand, London WC2R 2LS, United Kingdom

○ CNR-Istituto Nanoscienze, Euromediterranean Center for Nanomaterial Modelling and Technology (ECMT), via Arnesano, I-73100 Lecce, Italy

■ Dipartimento di Matematica e Fisica "Ennio De Giorgi", Università del Salento, via Arnesano I-73100 Lecce, Italy

\* E-mail: andrea.camposeo@cnr.it

\* E-mail: riccardo.sapienza@kcl.ac.uk

\* E-mail: dario.pisignano@unisalento.it

ABSTRACT: Nanoscale generation of individual photons in confined geometries is an exciting research field aiming at exploiting localised electromagnetic fields for light manipulation. One of the outstanding challenges of photonic systems combining emitters with nanostructured media is the selective channelling of photons emitted by embedded sources into specific optical modes and their transport at distant locations in integrated systems. Here we show that soft-matter nanofibres, electrospun with embedded emitters, combine subwavelength field localisation and large broadband near-field coupling with low propagation losses. By momentum spectroscopy we quantify the modal coupling efficiency identifying the regime of single-mode coupling. These nanofibres do not rely on resonant interactions making them ideal for room-temperature operation, and offer a scalable platform for future quantum information technology.

Light-matter interaction design in nano-structured optical environments has revolutionised the nanoscale control of light emission, absorption, and transport. Exerting such control on single-photons generated by individual emitters is also a fundamental step towards integrated quantum technologies, with applications including single-photon transistors,[1,2] on-chip generation and routing of single photons,[3-5] quantum interference,[6,7] or long-range coupling of single emitters.[8]

In this framework, plasmonic systems, which exploit near-field interactions in narrow gaps or nano-sized resonators,[9,10] have attracted much attention for broad-band and room temperature operation: plasmonic nano-antennas can increase the directionality[11] and decay-rate[12,13] of single-photon sources while the emitted photons can be transported to distant locations by plasmon modes in metal nanowires,[14] wedges,[15] or V-groove metal surfaces.[16]

Inspired by plasmonic architectures, subwavelength dielectric nanostructures are emerging as candidate materials for light management at room-temperature. Unlike photonic crystals, which exploit long-range interference (over many wavelengths) to design narrow resonance modes, subwavelength dielectrics offer practical local field enhancements and the ability of near-field light manipulation with the advantage of being free of the metal optical losses.[17-19]

Indeed, these waveguides can channel and transport light from a single emitter across long distances with relatively low losses, realising a simple interface between light emitters and propagating modes,[20] increasing off-chip extraction efficiency with diamond nanowires[21] and semiconductor nanopillars.[22,23] Moreover, the interaction of an emitter with a subwavelength waveguide allows for the exploitation of unique phenomena, such as spin-orbit coupling and unidirectional emission which occur only when light is spatially confined.[24, 25]

Polymer subwavelength waveguides fabricated by self-assembly, drawn from polymer solutions and melts,[26] are a promising system compatible with integration of quantum dots or molecules as emitters, and with potential for scalability and interconnection to form networks in 2-dimension[27] and 3-dimensions.[28] This improves on conventional pulled silica fibres with emitters evanescently coupled on the nanofibre surface[5, 29] and on pulled capillary hosting single molecules.[30] Moreover, these waveguides are cavity-free and with a broadband response, which makes them ideal for room-temperature (where sources are broadband) and ultrafast operation and for further interfacing with biological matter.

While polymer nanofibres do not require nano-lithography or top-down nanofabrication to be realized, the emitter-waveguide coupling has to be assessed *a posteriori*. In this respect, characterising the different modal contributions is important for further optical engineering of the propagation and coupling properties, and for quantum applications. The coupling efficiency can be probed from the collected scattered and transmitted light when the waveguide terminals are accessible,[5,29] by measuring the excitation extinction,[30] or when coupling to strong resonances from indirect estimations, such as those drawn from the lifetime variations of on- and off-coupling.[18,31] To date, quantifying the coupling efficiency of a single emitter to a specific nanofibre mode, in a way valid also for subwavelength and weakly resonant modes, is an outstanding challenge.

Here we report broadband single-photon generation and transport from isolated quantum dots embedded in the core of a free-standing and subwavelength polymer nanofibre. By means of momentum spectroscopy, we record the full *k*-vector distribution of the light emitted by individual quantum dots, revealing the selective coupling of the different dipole orientations to the individual nanofibre modes. We find a coupling to the fundamental mode with values up to $\beta_{01}= (31\pm2)\%$ which compare well with the theoretical maximum coupling of $\beta_{01}= 53\%$.

## RESULTS AND DISCUSSION

The hybrid nanofibre-emitter system is designed for (i) maximal coupling, by embedding the emitter at the nanofibre core, and (ii) efficient long-range transport, which is achieved by a free-standing geometry. The nanofibres are fabricated by electrospinning a polymethylmetacrylate (PMMA) solution (see Methods) doped with CdSeTe colloidal quantum dots (Invitrogen QD800, emission peaked at 790 nm). Unlike electron-beam or focussed ion-beam lithography, the electrospinning process largely preserves the integrity of embedded organic and inorganic light emitters.[26] Doped fibres with a diameter (D) between 300 nm and 1000 nm, depending on the processing parameters, are deposited either (i) on a glass substrate or (ii) on a transmission electron microscopy (TEM) grid producing free-standing nanostructures, with circular cross-section and length of tens to hundreds of micrometres with isolated quantum dots. A scanning electron microscopy (SEM) image of a typical free-standing nanofibre is shown in Figure 1a. When illuminated with a red laser (632 nm), the emitters can be located *via* wide-field fluorescence imaging, as in Figure 1b. The emission dynamics is studied by confocal microscopy (numerical aperture, NA = 0.95) and time correlated single-photon counting (TCSPC). Hanbury-Brown and Twiss interferometry of the quantum dots in free-standing nanofibres using two avalanche photodiode (APD) in start-stop configuration is shown in Figure 1c. The intensity correlation function $g_2(t)$ measured under continuous wave (CW) excitation (power 12 μW, wavelength 633 nm) is fitted to $g_2(t) = A(1 - \frac{1}{N} exp(|t|/\tau'))$, which gives a second-order intensity correlation function at zero-time, $g_2(0) = 0.10 \pm 0.03$, and a corresponding decay-time $\tau' = (84 \pm 3)$ ns, which confirms single-photon emission. The excited state lifetime for the quantum dot is measured under picosecond pulsed excitation (average power 1.5 μW, repetition rate 2.5 MHz, wavelength 634 nm). In the specific case shown in Figure 1d, a single exponential fit gives a decay time $\tau$ = (137 ± 0.1) ns. The difference between $\tau$ and $\tau'$ can be explained *via* an incoherent repumping term R, with $\tau' = (1/\tau + R)^{-1}$, as in Ref. 32. Fluorescence lifetime statistics of individual quantum dots inside free-standing nanofibres show a decay rate reduction of ~25% as compared to a reference homogenous polymer film, in agreement with the theoretical prediction (Supporting Figure S1). The decay rate of an emitter inside a subwavelength nanofibre is expected to be weakly reduced, due to the low refractive index of PMMA (nPMMA = 1.49, see Supporting Figure S1). Light collected from the nanofibre-coupled quantum dots has a typical blinking behaviour with an off state of 300 counts/s (Figure 1e).

For a quantum dot in the nanofiber, the quantum dot saturation intensity is 7 kWcm-2, achieved at an average pump intensity of 0.5 µW in a diffraction-limited illumination area (air objective, NA=0.95), for which we have measured 38 kphotons/s emitted, with an expected maximum count-rate (S∞) of 42 kphotons/s (see Supporting Figure S2). Above saturation the quantum dot approaches one photon emitted per excitation pulse, *i.e.* 2.5 Mphotons/s, given our laser repetition rate of 2.5 MHz, limited by the less than unitary quantum efficiency (~0.7, as provided by the manufacturer, Invitrogen) and by the off-states of the quantum dot (~30% on-state due to blinking) to a value of ~0.5 Mphotons/s.

Momentum spectroscopy of the emitted light directly accesses the coupling of the emitter to the nanofibre modes and allows us to calculate the modal coupling in an original way, without measuring the light transported by the nanofibre. Here we apply Fourier patterns analysis which is emerging as a reliable quantitative tool to probe otherwise inaccessible information about the emitter [33] and its emission directionality.[11] We measure the light emitted in momentum space by recording the angular patterns of the radiation emerging from a quantum dot inside a nanofibre lying on glass. Index matching enables access to large angles, up to the 1.45 numerical aperture (NA) of our objective, which encompasses the wave-vectors of the guided modes beyond the air light-line. We collect Fourier patterns as shown in Figure 2a and 2b for two nanofibres: a thin one, in the single mode regime ($D/\lambda = 0.5$, Figure 2a) and a thick one, in the multimode regime ($D/\lambda = 0.9$, Figure 2b). Each pair of momenta lobes (positive and negative kx where x is the direction of the longitudinal axis of the nanofibre) beyond the air light-line ($|k_x| > 1$), elongated in a direction orthogonal to the nanofibre, is the k-vector distribution of the distinct nanofibre modes. This is a clear experimental signature of the individual coupling of the quantum dot to each of the nanofibre modes which is in very good agreement with the calculated Fourier patterns shown in Figure 2c and 2d.

Figure 2e plots the mode dispersion structure reconstructed from the calculated angular pattern of a dipolar source in the centre of a nanofibre (see Methods). The intensity map indicates the coupling strength of a longitudinal (blue) and transverse (red) dipole to the nanofibre, normalised to unity, which is the maximal value in the plot. The guided modes are visible as two lines that stem from the air light-line $\omega = c_0 k_x$ growing to larger k-vectors up to the polymer light-line $\omega = c_0 k_x / n$, where $\omega$ is the light frequency, $c_0$ the speed of light in vacuum and n the refractive index. Due to the different spatial symmetry of the modes, the transverse (red) dipole couples almost exclusively to the fundamental mode $LP_{01}$, and the longitudinal (blue) to the second higher mode $LP_{11}$. Instead, the light emitted inside the light cone in air ($\omega > c_0 k_x$) corresponds to uncoupled radiation, emitted into free-space modes. We confirm this selective coupling and the nature of the modes by superimposing the analytical solution (dashed white lines) of the dispersion relations of the modes for a free-standing nanofibre.34 The analytical solution matches the maxima of the coupling of the two dipole-orientations which confirms that the strength of the momenta peak is a direct measure of the coupling of the emitter to a specific wave-vector, *i.e.* to a specific nanofibre mode.

Coupling of an emitter to the propagating waveguide modes is required for transport and manipulation of the emitted single photons. Plasmonic nanostructures can offer very high coupling efficiencies, as high as 21% for silver nanowires[14] or 42% for V-groove channels[16], although propagation of coupled photons is limited by the strong Ohmic losses in the metal. Instead dielectrics can offer much longer propagation length at the price of a weaker coupling, which has been measured at room-temperature to reach ~30% for quantum dots outside a silica nanofibre, approaching experimentally[5,29] the theoretical limit imposed by evanescence coupling.[35] Higher values, close to 100%, can be reached at cryogenic conditions for the coupling of epitaxial quantum dots to photonic crystal waveguides[31] although this is limited to narrow-band resonance, or ~18% for single molecules inside a broadband pulled glass capillary.[30]

Here we developed a different technique based on the momentum spectroscopy of the emitted light; from the ratio of the coupled and uncoupled light the total coupling constant $\beta$ as well as the coupling to the fundamental mode $\beta_{01}$ can be estimated. The coupling efficiency $\beta$ is obtained as the ratio between the light emitted into the guided modes region ($|k_x|>1$) and the total emitted light $\beta = I(|k_x| > 1)/I$, while $\beta^{01}$ is the ratio between the intensity of the k-vector corresponding to the mode $LP_{01}$ and the total emitted light (see Methods). The obtained coupling values are shown in Figure 3, where the lines represent the theoretical predictions and the points indicate the experimental data. While the total coupling $\beta$ remains roughly constant for all diameters (see Supporting Figure S3), the theoretical coupling $\beta_{01}$ to the fundamental mode (red line in Figure 3) reaches its maximum value of $\beta_{01} = 53\%$ at $D/\lambda = 0.55$, and then drops for larger diameters as more modes become available. The measured coupling values (green bars in Figure 3) are in the range $\beta_{01} = (21 \pm 4)\% - (31 \pm 2)\%$ for single mode nanofibres, and $\beta_{01} = (5 \pm 1)\% - (15 \pm 1)\%$ for multimode nanofibres, where also the coupling to the second higher mode $LP_{11}$ increases (orange bars in Figure 3). The errors come from the uncertainty in the position of the peaks in the k-space. The accuracy of this estimation can be confirmed by FDTD calculations that show an agreement within 10% between the values of β obtained from the two methods of momentum spectroscopy and direct modal coupling (see Supporting Information and Figure S4). As expected, the experimental values are lower than the theoretical prediction for a transverse or longitudinal dipole, as in the experiments the quantum dot's dipole moment is randomly oriented. These findings confirm that large single mode coupling can only be achieved with subwavelength nanofibres. The momentum spectroscopy method here developed is also ideal for probing nanofibres with more complex modes, as for example periodically corrugated nanofibres,[36] where the induced bandgaps would be evident in the dispersion relation.

When the nanofibre is intentionally doped with many quantum dots, their fluorescence at location distant from the illumination spot is a direct evidence of transport of the exciting laser light through the nanofibre, as shown in Figure 4a which shows quantum dots excited as far as 60 μm from the laser spot. The granularity of the quantum dot doping prevents a quantitative estimation of the propagation length, which instead can be done by recording how the fluorescence intensity leaks from a nanofibres homogeneously doped with a near-infrared dye, as shown in Figure 4b. By fitting with an exponential model the leaked light we obtain an optical loss coefficient $\xi^{-1} = (105 \pm 5)\ cm^{-1}$, corresponding to a propagation length $\xi = (95 \pm 5)\ \mu m$. Both the nanofibre scattering and the dye re-absorption (~ 58 cm-1, from bulk absorption measurements) are involved in determining the resulting $\xi$ value. Theoretically, for a nanofibre in air we expect a propagation length of 100-1000 µm, as calculated by including Rayleigh scattering from the nanofibre surface roughness (few nm)[37] as measured by atomic force microscopy.

As both long propagation and large coupling are important criteria to assess the quality of an hybrid nanoscale architecture, the most recent figure of merit (FOM) introduced by Bermúdez-Ureña et al.[16] combines them in the product $(\beta\, P\, \xi)\,/\lambda$, of mode coupling ($\beta$), Purcell factor ($P$) and propagation length ($\xi$) normalised by free-space operating wavelength ($\lambda$). We suggest here that $\beta$ is replaced by the modal coupling, i.e. $\beta_{01}$ for the fundamental mode, which has more practical importance for single-mode operation. The state-of-the-art in plasmonic systems is reported to be FOM = 6.6 ± 23%, while for our dielectric nanofibres we can calculate a FOM of 44 ± 16% coming from a coupling of $\beta_{01} = (31 \pm 2)\%$ and a Purcell factor with respect to vacuum of $1.2 \pm 15\%$ and the propagation length $\xi = (95 \pm 5)\,\mu m$. While plasmonic systems can strongly engineer the Purcell factor, dielectric nanostructures, with negligible Ohmic losses, offer longer propagation lengths, resulting in a large FOM. This FOM is useful to compare nanoscale geometries, while it loses importance for macroscopic systems such as conventional single mode fibres for which the very long propagation length dominates the FOM.

## CONCLUSIONS

In conclusion, we report on a nanofibre single-photon light source architecture, integrating a quantum emitter into a low loss single mode optical waveguide, and operating at room temperature. Using an approach based on momentum spectroscopy, we have quantified the coupling efficiency of individual quantum dots to each nanofibre mode, obtaining broadband coupling of up to 31% of the emitted light (see also Supporting Figure S5) and a FOM of $44 \pm 16\%$. Through their robustness and flexibility, these nanofibres have potential as a future platform for nanoscale quantum optics as they can be connected to form a network of coupled emitters for quantum technological applications. Moreover, these systems are compatible with plasmonic systems, and therefore suitable for further photonic engineering.

METHODS

*Nanofibres fabrication.* 1 nM solution containing CdSeTe quantum dots (Invitrogen QD800, emitting at 790 nm) was mixed with a chloroform solution of PMMA (MW 120000, Sigma Aldrich). In order to obtain sub-micron fibres, 30 mg of tetrabutylammonium iodide (Sigma Aldrich) organic salt was added to the solution. By measuring the emission spectrum of the quantum dots, we checked that the solvent, polymer matrix and organic salt do not affect the photoluminescence properties of the embedded emitters. The background fluorescence from PMMA is estimated to be <1% of the emission from the QD, as determined both by measuring the background fluorescence collected from the excited spot and by exciting 10 μm away from the edge of an intentionally cut nanofiber and collecting at the edge (See Supporting Figure S5). For light transport measurements, the IR-140 organic chromophore was added to the PMMA solution, using a concentration of 0.1% weight:weight relative to the PMMA polymer matrix. The solution was loaded in a 1 mL syringe tipped with a 27G gauge stainless steel needle and delivered at 0.5 mL h$^{-1}$ constant rate by means of a syringe pump (Harvard Apparatus, Holliston, MA). A bias voltage of 8 kV was applied between the needle and the metallic collector by a high voltage power supply (Glassman High Voltage). Glass coverslip substrates (18×18 mm$^2$) were placed on a 10×10 cm$^2$ Cu plate used as grounded collector, positioned 10 cm away from the needle. Free-standing nanofibres were obtained by deposition on a TEM grid (TAAB Laboratories Equipment Ltd) used as collector. Electrospinning was performed with a relative humidity and temperature of about 40% and 25 °C, respectively. Reference thin films on glass coverslip substrates were obtained by spin-coating the solution used for electrospinning at 6000 rpm for 40 s. The nanofibre morphology was investigated by SEM (FEI Nova NanoSEM 450 or Zeiss) and atomic force microscopy (Veeco Instruments). The fibre diameters were in the range of 300-1000 nm, with a surface roughness (root-mean-square, RMS) of the order of few nm. The fibres have almost round-shape cross-section (Fig. S6a,b), with a ratio between the height and the width in the range 0.9-1. A maximum variation of 70 nm of the fibre diameter over a length of 1 mm was measured (Fig. S6c-h).

*Confocal optical microscopy.* The excited state lifetime of isolated quantum dots was measured by confocal microscopy and TCSPC (TimeHarp 260, PicoQuant) with an overall temporal resolution of ~100-200 ps. The quantum dots were excited with a 100 ps, 634 nm, 2.5 MHz repetition rate laser (average power 1.5 µW) through an air objective (NA = 0.95) or an oil immersion objective (NA=1.45). The light emitted was collected by the same objective and directed to an avalanche photodiode (APD). The lifetime traces were fitted with both a single exponential and a lognormal model, and the one providing the highest accuracy was retained. Antibunching was measured through the same setup, but using two APDs in start-stop configuration. The intensity correlation function $g_2(t)$ measured from a quantum dot under CW excitation was fitted to $g_2(t) = A(1 - \frac{1}{N} exp(|t|/\tau'))$. The collection efficiency of our microscope is estimated to be ~5% for polymer films and 1.45 NA oil immersion objective and ~2-3% for free-standing fibres and 0.95 NA air objective. These values are obtained as the ratios between the measured saturation intensity of several individual QDs and the expected emission intensity at saturation (laser repetition rate times the quantum efficiency)[5].

*Optical Fourier imaging.* We recorded the angular emission patterns $I(k_x, k_y)$ of isolated quantum dots by imaging the back-focal plane into a CCD camera (Princeton Instruments, Pixis 400). The excitation and collection was achieved through an oil immersion objective (100×, NA=1.45). We calculated the intensity distribution $I(k_x) = \sum_{k_y} I(k_x, k_y)$, where $k_x$ is the component of $\vec{k}$ along the nanofibre axis. We estimated the nanofibre diameter $D$ by comparing the position of the peaks in the experimental Fourier patterns with the theoretical modal dispersion. As the nanofibre modes are only marginally affected by the glass interface, the light emitted in the *k*-regions beyond the light-cone in air corresponds to guided modes ($|k_x| > 1$), and the light inside the light-cone in air to the uncoupled modes. For multimodal fibres, owing to the modes' orthogonality in *k*-space, different modes have different wave-vectors for the same energy; this allows assessing the individual coupling to these modes by measuring the relative strength of the Fourier peaks. In this way, the coupling to the fundamental mode $\beta_{01}$ can be separated from the total coupling β to all the waveguide modes.

The coupling efficiency $\beta$ is obtained as the ratio between the light emitted into the guided modes region ($|k_x|>1$) and the total emitted light:

$\beta = I(|k_x| > 1)/I,$

while $\beta_{01}$ is the ratio between the intensity of the *k*-vector corresponding to the mode LP$_{01}$ and the total emitted light. This can be generalised for all other higher-energy modes.

An uncertainty is attributed to both *D* and β due to the uncertainty in the *k*-space calibration that we quantify as 0.05. The accuracy of this estimation can be confirmed by Finite Difference Time Domain (FDTD) calculations that show an agreement within 10% between the values of β obtained from the two methods of momentum spectroscopy and direct modal coupling (See Supporting Figure S4).

*Numerical calculations*. FDTD simulations were performed using a commercial package (Lumerical). The local density of optical states (LDOS) maps were obtained by considering the intensity emitted by a dipolar source in different position inside the nanofibre. The total coupling efficiency was obtained by calculating the intensity transmitted through a monitor crossing the free-standing nanofibre 4 µm away from the source; the modal coupling efficiency was obtained by mode projection of the same intensity onto the independently calculated nanofibres modes. The angular patterns were obtained by means of far field projections of the electro-magnetic fields in a plane 300 nm outside the nanofibre inside the substrate of a nanofibre laying on glass. Calculations with and without the glass substrate show that the low-refractive index interface below the nanofibre only weakly affects the modal dispersion, inducing only a spectral broadening due to the increased losses into the glass.

ASSOCIATED CONTENT

**Supporting Information Available**:

Purcell factor measurement, data on quantum dot saturation intensity, additional calculation of modal coupling and information about momentum spectroscopy, data about broad band coupling of quantum dots to nanofibres and atomic force microscopy. This material is available free of charge *via* the Internet at http://pubs.acs.org

Supporting Information (PDF)

AUTHOR INFORMATION

**Corresponding Authors**

* riccardo.sapienza@kcl.ac.uk, * andrea.camposeo@nano.cnr.it, * dario.pisignano@unisalento.it

**Notes**

The authors declare no competing financial interest.

ACKNOWLEDGMENT

We wish to thank Simon Fairclough for fruitful discussions. The research leading to these results has received funding from the Engineering and Physical Sciences Research Council (EPSRC), from the European Union's Seventh Framework Programme (FP7/2007-2013) EU Project People, from the Leverhulme Trust (RPG-2014-238), from the Royal Society (Research Grant n. RG140457), and from the European Research Council under the European Union's FP/2007-2013/ERC Grant Agreement n. 306357 ("NANO-JETS"). The data is publicly available in Figshare[38].

Published in ACS Nano, doi: [10.1021/acsnano.6b02057](10.1021/acsnano.6b02057) (2016).

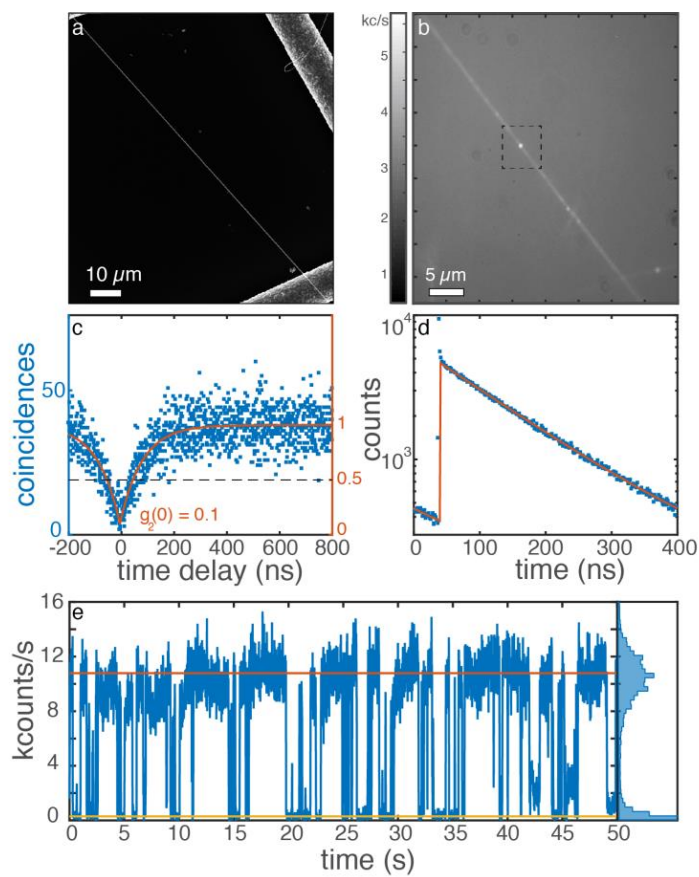

**Figure 1.** Addressing a single quantum dot coupled to a nanofibre. (a) SEM picture of a free-standing nanofibre. (b) Wide-field fluorescence image of a similar nanofibre highlighting the presence of isolated quantum dots (dashed square). (c) The discrete nature of the photons emitted by the single quantum dot is evidenced by the second order correlation function, obtained via continuous-wave laser excitation (12 µW, 633 nm) which leads to $g_2(0) = 0.1 \pm 0.03$ when fitted with $g_2(t) = A(1 - \frac{1}{N} exp(|t|/\tau'))$. (d) The fluorescence dynamics of the same quantum dot under picosecond excitation (1.5 µW, 2.5 MHz, 634 nm) is fitted with a single exponential decay leading to an excited state lifetime $\tau = (137\pm0.1)$ ns. (e) The time trace of the photons emitted by the quantum dot shows a blinking behaviour with on- (~10.5 kcounts/s) and off- (~0.3 kcounts/s) states as highlighted in the histogram.

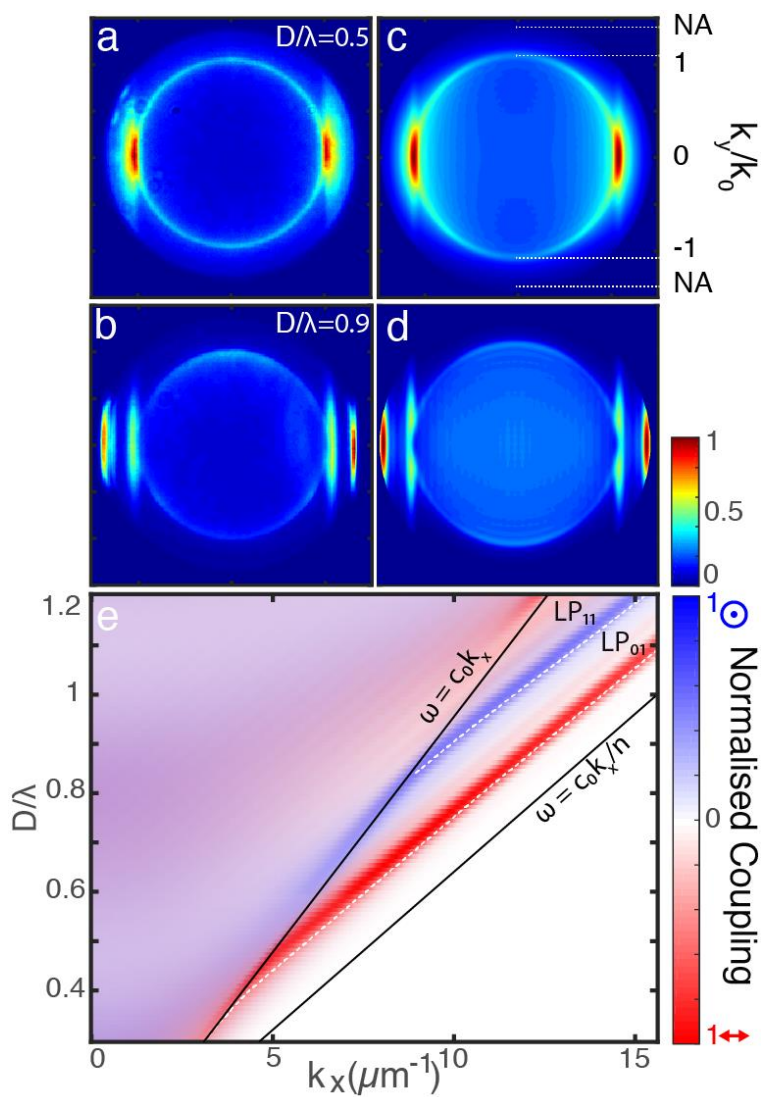

**Figure 2.** Momentum spectroscopy analysis. (a) and (b) Experimental angular patterns and, (c) and (d), the corresponding theoretical calculations (FDTD) of the radiation emitted from a quantum dot in nanofibres on glass of different diameter for $D/\lambda = 0.5$ (a-c) for $D/\lambda = 0.9$ (b-d); $k_0$ is the light wavevector in air. (e), theoretical calculations of the coupling of a dipole to the modes of the nanofibre obtained from the angular emission patterns of a longitudinal (blue) and transverse (red) dipole located at the centre of the nanofibre, normalized to the maximum coupling (see Methods). The light-line in air and in the polymer is indicated by the black lines, delimiting the region with guided modes. Inside the air light-cone, the light corresponds to uncoupled emission and is therefore similar for both dipolar orientations. Instead, the emission for *k*-vectors beyond the air light line can couple to the nanofibre modes (high intensity red and blue bands). The emission maxima correspond to the dispersion of the two nanofibre modes, proving that for reasons of symmetry the transverse dipole couples exclusively to the $LP_{01}$, and the longitudinal dipole to $LP_{11}$. The analytical dispersion relation of the first two modes of the nanofibre in air ($LP_{01}$ and $LP_{11}$) are plotted as white dotted lines.

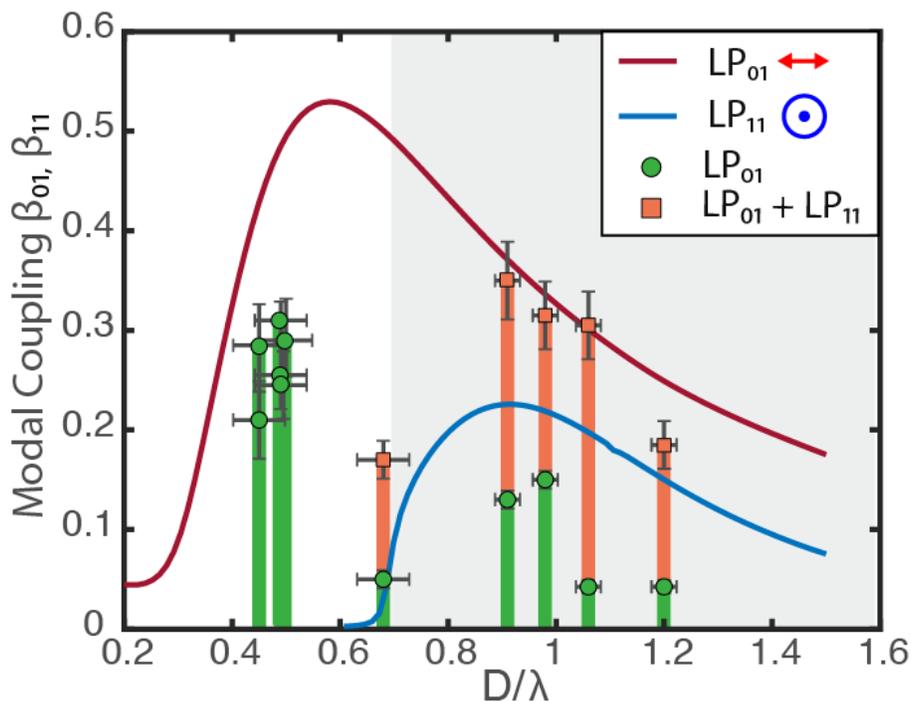

**Figure 3.** Modal coupling of the quantum dot to the nanofibre. Estimation of the emitter-nanofibre coupling via momentum spectroscopy. The experimental values for the coupling to the fundamental mode $LP_{01}$ are plotted as circular green points (and green bar). In the multimode region (grey shaded area), coupling to the second higher mode $LP_{11}$ are shown by an orange bar and the total coupling is marked by orange squares. The experimental values are compared to the theoretical calculations for the two dipole-orientations (red line for transverse, and blue line for longitudinal). The coupling to the fundamental mode $\beta_{01}$ quickly drops for $D \geq 0.7\lambda$ due to the emergence of other modes. The error bars originate from the uncertainty in the position of the peaks in the *k*-space.

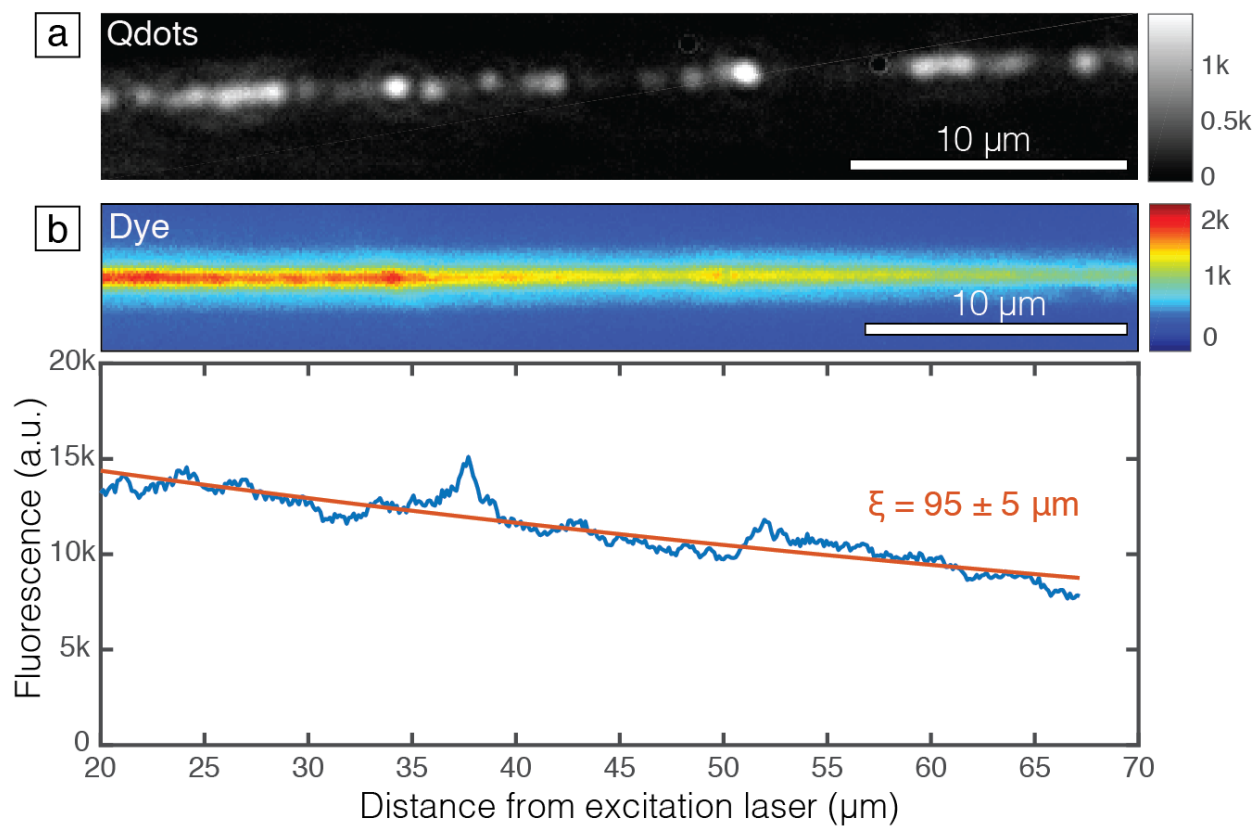

**Figure 4.** Light transport through the nanofibre. Panel (a) shows a nanofibre with an intentional high-density of quantum dots when remotely excited. The excitation position is outside the field of view. The laser beam couples to the nanofibre and is transported through it, which is revealed by the quantum dots excited distant locations. (b) A more quantitative measure of the transport length can be obtained by exciting a nanofibre doped with an IR dye emitting around 8900 nm. The excitation is 20 μm outside the field of view. The fluorescence intensity recorded along the nanofibre is fitted with and an exponential function [$A\,exp(-x/\xi)$] after subtracting the background, obtaining a propagation length of $\xi = (95\pm5)$ μm.

# Supporting Information

# Modal Coupling of Single Photon Emitters Within Nanofibre Waveguides

*Michele Gaio*[◆], *Maria Moffa*[○], *Marta Castro-Lopez*[◆], *Dario Pisignano*[○,■]*, *Andrea Camposeo*[○]*, *and Riccardo Sapienza*[◆]*

◆ Department of Physics, King's College London, Strand, London WC2R 2LS, United Kingdom

○ CNR-Istituto Nanoscienze, Euromediterranean Center for Nanomaterial Modelling and Technology (ECMT), via Arnesano, I-73100 Lecce, Italy

■ Dipartimento di Matematica e Fisica "Ennio De Giorgi", Università del Salento, via Arnesano I-73100 Lecce, Italy

## Purcell studies

Light sources placed inside a subwavelength nanofibre suffer from suppressed emission rates due to a smaller than unity Purcell factor at the high-low index (PMMA-air) interface coming from the reduced local density of optical states (LDOS) [S1]. We measured the decay rate $\Gamma$ distribution of individual quantum dots inside free-standing nanofibres as shown in Figure 1d, and compared them to a reference PMMA film. While both systems show a large $\Gamma$ distribution with a full width at half maximum around ~50%, when the emitters are inside the free-standing nanofibre the average decay rate decreases from $12 \times 10^{-3}$ ns$^{-1}$ to $9 \times 10^{-3}$ ns$^{-1}$ (Figure S1a). Assuming that the quantum dots dipoles randomly oriented in the nanofibre, the data agrees well with the theoretical predictions that we obtained through the finite difference time domain (FDTD) method obtained by considering the intensity emitted by a dipolar source in different position inside the nanofibre, plotted in Figure S1b and S1c.

Figure S1b and S1c show the LDOS maps normalized to the bulk material for two dipolar orientations inside the PMMA sub-wavelength nanofiber (D = 300 nm). The Purcell factor is close to unity (in the range 0.8 to 1) in the case of a longitudinal dipole (Figure S1b) and conversely, it is reduced to around 0.5 for a transverse dipole (Figure S1c). For a transverse dipole in a subwavelength Silicon fibre (n = 3.4, D =130nm) the Purcell factor is ~0.1.

## Quantum dot saturation

Figure S2 shows the emitted intensity from a single quantum dot embedded in a free-standing nanofibre as a function of the illumination power. A typical saturation curve is, with a saturation at an average pump intensity of ~0.5 µW, where we collect 25 kphotons/s from the nanofibre. Well above saturation, the maximum measured count-rate is 38 kphotons/s, when the dot emission is expected to approach 2.5 Mphotons/s, given our laser repetition rate of 2.5 MHz, a value which is limited by the less than unitary quantum efficiency (~0.7) and by the off-state of the quantum dot (~ 30% on-state due to blinking) to a value of ~0.5 Mphotons/s. The line in Figure S2 is a fit to the data using the saturation curve of a two-level system: $S = S_\infty (I/I_s)(1 + I/I_s)^{-1} [1 - \exp\{-(1 + I/I_s) \tau_p/\tau_f\}]$ where I and $I_s$ are respectively the excitation and the saturation intensities, respectively, and $\tau_p$ the excitation laser pulse width (100 ps) and $\tau_f$ the decay time of the quantum dots [S2]. From the fit we obtain a saturation count rate $S_\infty$ = 42 kphotons/s and $I_s$ = 7 KWcm$^{-2}$.

## Modal coupling calculations

The total and modal coupling of a quantum dot to the nanofibre modes is calculated by FDTD simulations using a commercial package (Lumerical). Figure S3 shows both the total coupling ($\beta$) and modal ($\beta_{01}$) to a specific fibre mode, the first excited one LP$_{11}$. The total coupling efficiency was obtained by calculating the intensity transmitted through a monitor crossing the free-standing nanofibre 4 μm away from the source; the modal coupling efficiency was obtained by mode projection of the same intensity onto the independently calculated nanofibres modes. The total coupling remains roughly constant for larger diameters, with oscillations due to the emergence of the different modes. Instead, the coupling to the fundamental mode $\beta_{01}$ peaks at $D/\lambda = 0.55$ and then quickly drops to zero for larger diameters, indicating that selective coupling to the fundamental mode is best achieved for subwavelength nanofibres.

## Momentum spectroscopy

We confirm numerically the accuracy of the estimation of the quantum dot - nanofibre coupling by momentum spectroscopy by FDTD calculations. Angular patterns were obtained by means of far field projections of the electro-magnetic fields in a plane 300 nm outside the nanofibre inside the substrate of a nanofibre laying on glass.

In Figure S4 the coupling to a nanofibre lying on glass calculated by far-field projections is compared to the value obtained by mode projection of the light transmitted in a free-standing nanofibre . The agreement is within 10% for both dipole orientations, for the considered diameter range D = 300-1000 nm. The discrepancy is explained due to the variable transmissivity of different $k$-components to the far field. For small diameters, the overestimation is also explained in terms of light emitted in the upper direction and not collected by the objective. In the real experiment, the collection is limited by the objective NA and response, which for large $k$-vectors further lowers the apparent coupling.

### Broad-band light coupling and transport

The nanofibre has a broadband response which encompasses the emission spectrum of the quantum dots. In Figure S5 we report the spectrum of the light emitted at the edge of an intentionally cut nanofibre once a distant quantum dot is excited. The quantum dot emission is well transported in its entirety, for a bandwidth of at least ~ 100 nm.

### Nanofibre morphology

The morphology of the realized PMMA nanofibers embedding quantum dots was investigated by atomic force microscopy (AFM). Supporting Figure S6a shows a typical AFM topographic image of a single nanofibre. The nanofibre has a circular cross section (ratio between the height

H and the diameter D, is H/D=0.96 in Fig. S6b). The AFM analysis performed on several fibers showed that H/D ratios are in the range 0.9-1, evidencing the uniformity of the cross-sectional shape of the produced fibers. Moreover, the variation of the nanofibre diameter over a length of few micrometers is of the order of 1% or less (Fig. S6c). To investigate this issue more in depth, we also measured the variation of the nanofibre diameter over a length of the order of 1 mm (Fig. S6d-h). A maximum variation of the diameter, $\Delta D$, of the order of 70 nm was measured over a length L=0.8 mm, providing a ratio $\Delta D/L=9\times10^{-5}$. Another important fiber morphological property for efficient light transport is related to the surface roughness. Indeed, Rayleigh scattering from surface defects is proportional to the variance of the surface roughness squared [S3]. AFM data provided a nanofibre surface roughness of about 3 nm, allowing for estimating a light transport length in the range 100-1000 µm.

## Supporting Figures

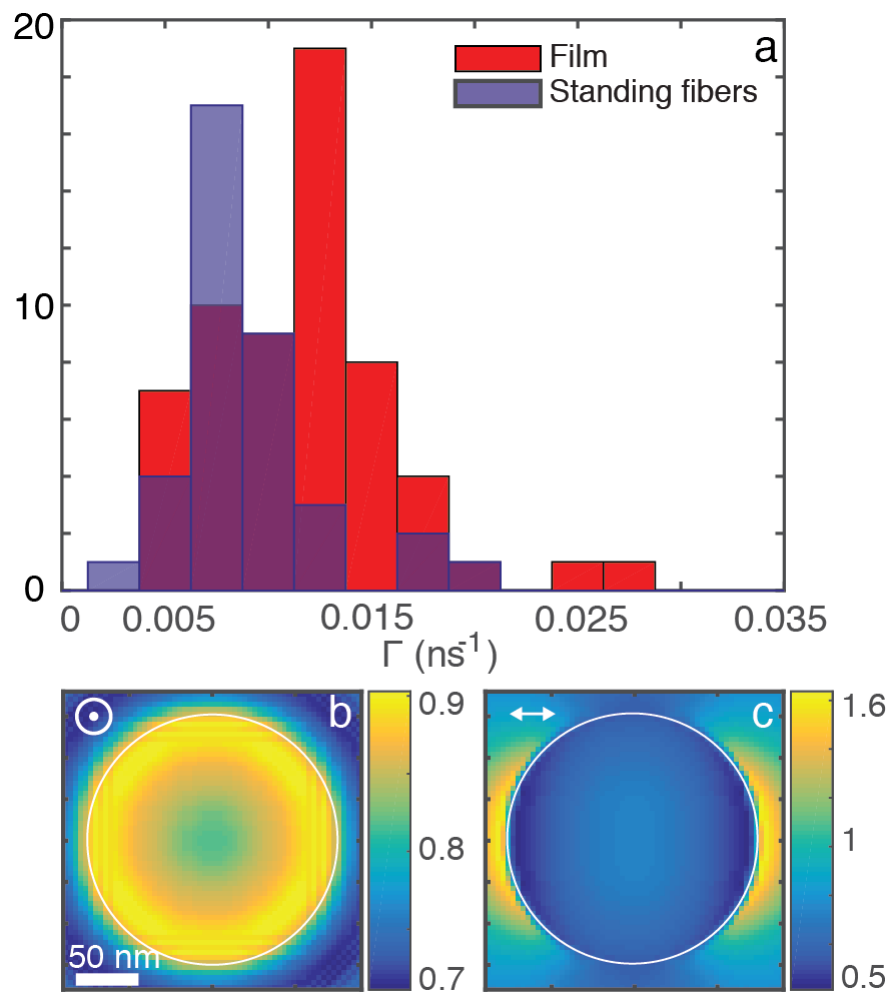

**Figure S1. Purcell effect for the coupled emitters. a,** Decay rates distributions measured from the quantum dots in the standing fibres (blue) and in a PMMA film used here as a reference. The shift of the decay rate distribution towards smaller values, from an average of $12\times10^{-3}$ ns$^{-1}$ to $9\times10^{-3}$ ns$^{-1}$, is compatible with the predicted values. **b** and **c,** Calculated LDOS maps, normalised to the value calculated for a bulk material, for a dipole longitudinal (**b**) and transverse (**c**) to the fibre with 300 nm diameter (*D*) at different transverse positions. For $D < \lambda$ the LDOS is almost uniform inside the nanofibre with values in the range of 0.45 to 1 depending on the dipole orientations. This is related to the presence of the interface that gets polarised by the dipole either in phase (longitudinal) or out-of-phase (transverse).

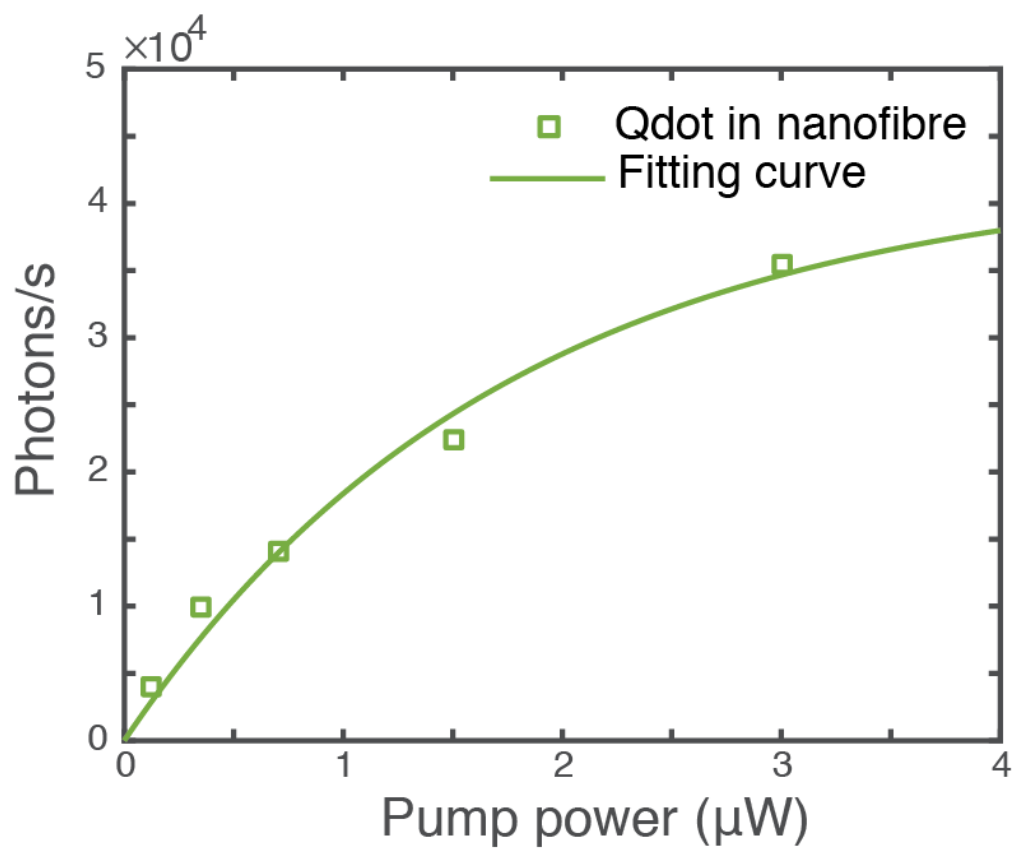

**Figure S2. Saturation**. Intensity emitted by a single quantum dot embedded in a nanofibre and in a reference film as a function of the illumination intensity. The repetition rate of our pulsed laser is set to 2.5 MHz. The quantum dot reaches saturation $S_\infty$ values of 42 kphotons/s in the nanofibre with a maximum measured count-rate of 38 kphotons/s, with a saturation intensity around 0.5 µW. The line is a fit to the data using a two level saturation model [S2].

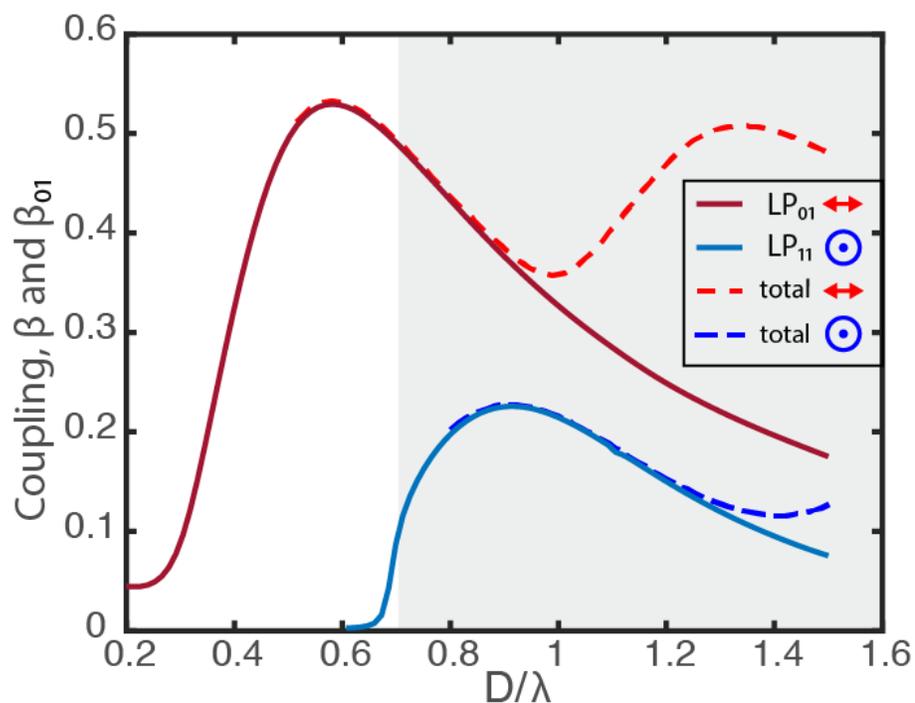

**Figure S3. Coupling of the quantum dot to the nanofibre**. Total ($\beta$) and modal ($\beta_{01}$) coupling is obtained by FDTD simulations (see Methods). While the total coupling (dashed lines) $\beta$ remains constant for larger diameters, the coupling to the fundamental mode $\beta_{01}$ (solid red line) and to the first mode (solid blue line) quickly drops for $D \geq \lambda$ due to the emergence of the other modes. The grey are indicates the multimode region.

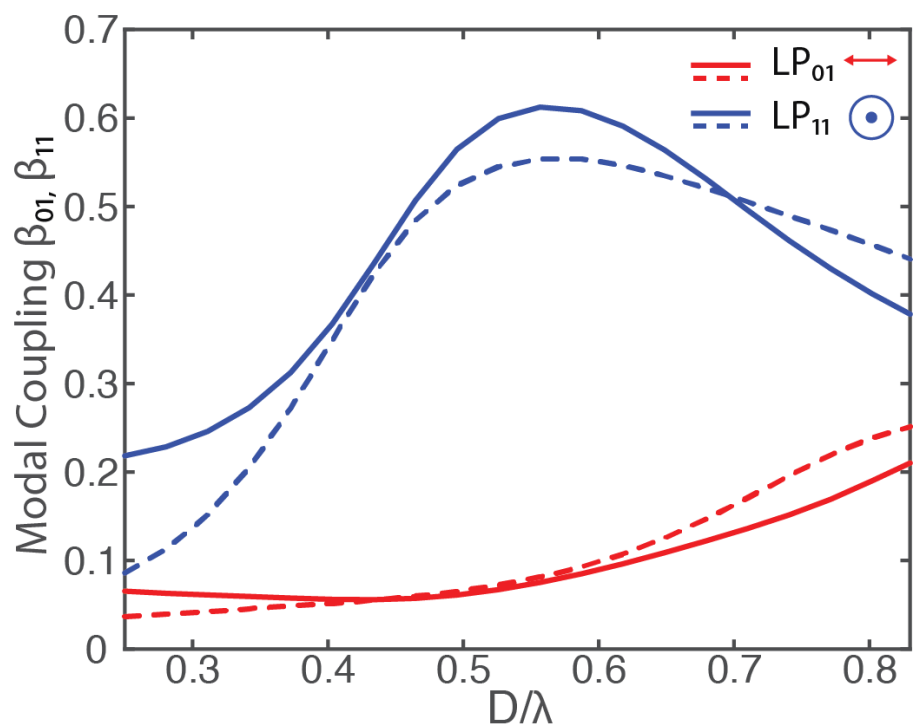

**Figure S4.** Coupling estimation (FDTD) from the momentum spectroscopy analysis (solid lines) of a nanofibre lying on glass compared to the direct probing of the transmission of a standing nanofibre (dashed line) for a longitudinally-oriented dipole (blue lines) and a transversally-oriented dipole (red lines). The comparison shows that the momentum spectroscopy analysis overestimates $\beta$ by a maximum of 10% for the diameter under examination.

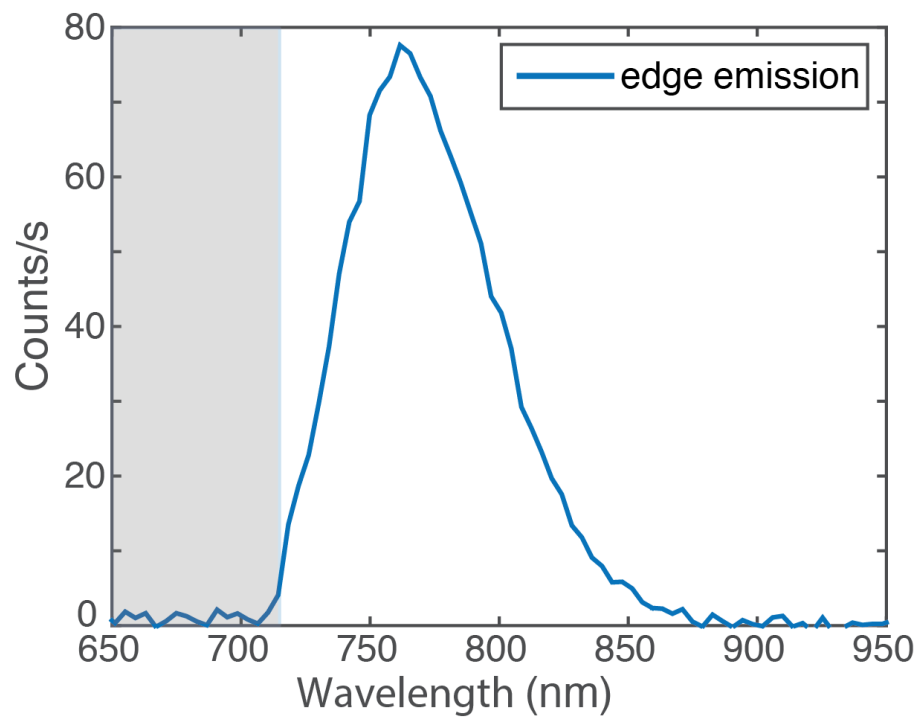

**Figure S5. Broadband character of the nanofibre-quantum dot system**. Measured spectrum of the light emitted at the edge of the intentionally cut nanofibre once a distant quantum dot is excited. The spectrum is transmitted in in its entirety as the nanofibre has a broad-band response. The grey area highlights the spectral rejection of the edge filter used to remove the excitation laser.

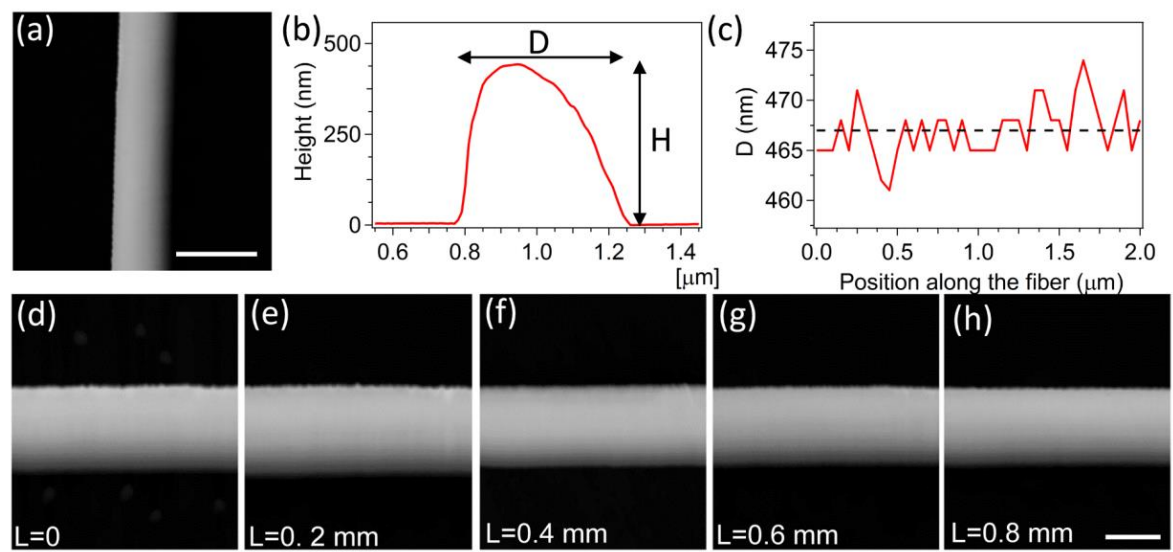

**Figure S6. AFM analysis**. (a) AFM image of the topography of a PMMA nanofibre embedding quantum dots. Scale bar: 500 nm. The corresponding horizontal height profile is shown in (b). The nanofibre has a diameter, D=460 nm and a height, H=440 nm. (c) Variation of the diameter, D, of the nanofibre shown in (a) along the nanofibre length. The dashed horizontal line highlights the average value of the nanofibre diameter ($D_{avg}$=467 nm, standard deviation: 3 nm). (d)-(h). AFM topographic micrographs measured in different points spaced 200 μm along the length (L) of an individual PMMA nanofibre. Scale bar: 300 nm.